\documentclass[aps,pra,showpacs,twocolumn]{revtex4-1}
\usepackage{amsmath}
\usepackage{amscd}
\usepackage{wasysym}
\usepackage[ansinew]{inputenc}
\usepackage[T1]{fontenc}
\usepackage{ae,aecompl}
\usepackage{amsfonts}
\usepackage{amsthm}
\usepackage{dsfont}
\usepackage{graphicx}
\usepackage{amsfonts}

\begin{document}
\title{Semiclassical quantisation for a bosonic atom-molecule conversion system}
\author{Eva-Maria Graefe, Maria Graney, and Alexander Rush}
\address{Department of Mathematics, Imperial College London, London, SW7 2AZ, United Kingdom}
\begin{abstract}
We consider a simple quantum model of atom-molecule conversion where bosonic atoms can combine into diatomic molecules and vice versa. The many-particle system can be expressed in terms of the generators a deformed $SU(2)$ algebra, and the mean-field dynamics takes place on a deformed version of the Bloch sphere, a teardrop shaped surface with a cusp singularity. We analyse the mean-field and many-particle correspondence, which shows typical features of quantum-classical correspondence. We demonstrate that semiclassical methods can be employed to recover full many-particle features from the mean-field description in cold atom systems with atom-molecule conversion, and derive an analytic expression for the many-particle density of states in the limit of large particle numbers. 
\end{abstract}
\maketitle

\section{Introduction}
The experimental progress in confining and manipulating cold atoms and Bose-Einstein Condensates (BECs) offers a unique opportunity to investigate the quantum properties of interacting many-particle systems. For most realistic setups, however, a theoretical full many-particle description is beyond the current state of the art. Most commonly the mean-field approximation is applied, resulting in a description of the many-particle system by an effective single particle wave function. The time evolution is in this description governed by a nonlinear Schr\"odinger equation. This approximation is closely related to the classical limit of single particle quantum systems, where the particle number plays the role of $\hbar^{-1}$. Recently there have been several studies demonstrating that this analogy can be used to apply semiclassical techniques to recover full many-particle features from the mean-field description alone \cite{07semiMP,Niss10,Chuc10,Visc11,Simo12,Simo14}. 

Over the last decade there has been considerable interest in atom-molecule conversion in cold atoms and BECs \cite{Vard01a,Sant06,Liu08,Li09,Liu10,Sant11,Khrip11,Li11b,Cui12,Shen13,Donl02}. Theoretically these systems are closely related to deformed $SU(M)$ algebras \cite{Bona96}. Neglecting quantum fluctuations in the many-particle dynamics leads to mean-field approximations defined on phase spaces with non-standard geometries, and new interesting features in the many-particle and mean-field correspondence. Here we address the question whether semiclassical methods might be adapted to describe the full many-particle behaviour of atom-molecule conversion systems on the grounds of their mean-field approximation. 

We focus on the simplest model of an atom-molecule conversion system, similar to the one considered in \cite{Vard01a}, consisting of non-interacting atoms and diatomic molecules each of which can populate only one mode. Introducing the atomic and molecular creation and annihilation operators,  $\hat a^\dag,\, \hat a$ and $\hat b^\dag,\, \hat b$, respectively, this system can be described by a Hamiltonian of the form 
\begin{equation}
\hat H=\epsilon_a \hat a^\dagger \hat a +\epsilon_b \hat b^\dagger \hat b+\frac{v}{2\sqrt{N}}(\hat a^\dag\hat a^\dag b+\hat a\hat a\hat b^\dag),
\end{equation}
where $\epsilon_{a,b}$ is the energy of the atomic or molecular mode, and $v$ describes the conversion strength between atoms and molecules. The total number of atoms $\hat N=\hat a^\dag \hat a +2\hat b^\dag \hat b$ is a constant of motion. For a fixed value of the particle number $N$ the system lives on an $\left[\frac{N}{2}\right]_<+1$ dimensional Hilbert space. For simplicity, we confine the discussions to even particle numbers in what follows. 

We begin with a review of the many-particle and mean-field descriptions, and their correspondence. Then we introduce a semiclassical quantisation condition and demonstrate that the many-particle spectrum can be accurately recovered from the mean-field dynamics, and derive an analytic expression for the many-particle density of states in the semiclassical limit of large particle numbers. We also present results for the semiclassical many-particle eigenstates. We end with a summary and outlook. 

\section{The many-particle system}
Similar to the two-mode Bose-Hubbard model \cite{Milb97}, we can apply a Schwinger-type transformation to introduce the operators
\begin{eqnarray} \nonumber
\hat{K}_x &=& \frac{\hat{a}^{\dag}\hat{a}^{\dag}\hat{b} + \hat{a}\hat{a}\hat{b}^{\dag}}{2\sqrt{N}},\\
\hat{K}_y &=&\frac{\hat{a}^{\dag}\hat{a}^{\dag}\hat{b} - \hat{a}\hat{a}\hat{b}^{\dag}}{2i\sqrt{N}},\label{SUoperators}\\ 
\nonumber
\hat{K}_z &=& \frac{ \hat{a}^{\dag}\hat{a}-2\hat{b}^{\dag}\hat{b}}{4}. 
\end{eqnarray}
The Hamiltonian can then be expressed as 
\begin{equation}
\label{eqn_MPHam}
\hat H=\epsilon \hat K_z+v\hat K_x,
\end{equation}
where we have shifted the zero energy and introduced the parameter $\epsilon=2\epsilon_a-\epsilon_b$.

\begin{figure}[tb]
\centering
\includegraphics[width=0.49\textwidth]{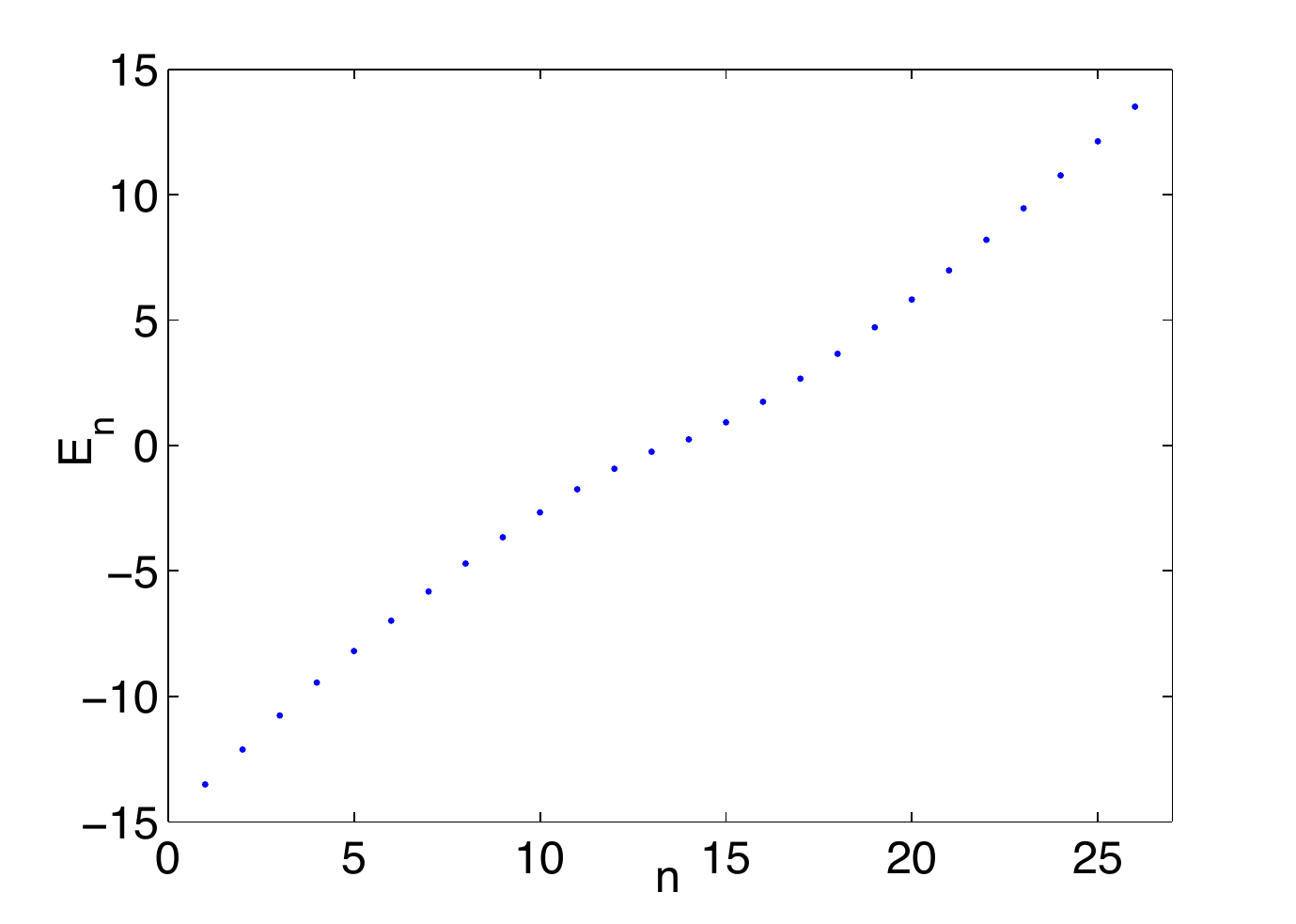} 
\caption{(Color online) Eigenvalues of $\hat K_x$ (and $\hat K_y$) for $N=50$.}
\label{fig_Kxspec}
\end{figure}

The physical meaning of the operators $\hat K_j$ is similar to the two-mode Bose-Hubbard case. That is, $\hat K_z$ measures the population imbalance between molecular and atomic mode, and $\hat K_{x,y}$ measure their phase relation. It is further convenient to introduce the operators 
\begin{equation}
\label{SUoperators_pm}
\hat K_\pm=\hat K_x\pm i\hat K_y,
\end{equation}
that create a molecule out of two atoms and vice versa. 
The operators (\ref{SUoperators}) and (\ref{SUoperators_pm}) are related to a \textit{nonlinear deformation} of an $SU(2)$ algebra  \cite{Bona96,Klim02c,Lee10}. That is, they fulfil commutation relations of the form
\begin{equation}
\label{eqn_defSU(2)_Kpm}
[\hat K_{z},\hat K_{\pm}]=\pm\hat K_{\pm},
\end{equation}
just as for $SU(2)$, and 
\begin{equation}
[\hat K_+,\hat K_-]=F(\hat K_z,\hat N),
\end{equation}
where $F(\hat K_z,\hat N)$ is a polynomial in $\hat K_z$ and $\hat N$. Specifically we have 
\begin{equation}
\label{poly_comm}
F(\hat K_z,\hat N)=-\frac{\hat N}{N}-\frac{1}{4N}\left(\hat N+4\hat K_z\right)\left(\hat N-12\hat K_z\right).
\end{equation}
In terms of $\hat K_x$, $\hat K_y$, and $\hat K_z$ the commutation relations read 
\begin{align}
[\hat K_{z},\hat K_x]&={\rm i}\hat K_{y},\label{eqn_defSU(2)_Kxy}\\
[\hat K_{y},\hat K_z]&= {\rm i}\hat K_{x},\label{eqn_defSU(2)_Kxyb}\\
[\hat K_x,\hat K_y]&=\frac{\rm i}{2}F(\hat K_z,\hat N).
\end{align}
The total particle number $\hat N$ commutes with $\hat K_x$, $\hat K_y$, and $\hat K_z$. In the case of a Bose-Hubbard dimer the conserved particle number reflects the conservation of the total angular momentum. For deformed $SU(2)$ algebras this is replaced by a less trivial conservation law. We can find the conserved quantity using the approach in \cite{Bona96} as
\begin{eqnarray} \nonumber
\hat C&=&\hat K_-\hat K_++\frac{4}{N}\hat K_z^3+\frac{\hat N+6}{N}\hat K_z^2+\frac{8-\hat N^2}{4N}\hat K_z \\
\nonumber&=& \hat{K}_x^2 + \hat{K}_y^2 + \frac{4}{N} \hat{K}_z^3 +\frac{\hat N}{N}\hat{K}_z^2 \\
&&+\frac{8-\hat N^2-4\hat N}{4N}\hat{K}_z+\frac{4\hat N+\hat N^2}{8N}.
\end{eqnarray}
Evaluating $\langle \hat C\rangle$ in any eigenstate of $\hat K_z$ yields the conservation law
\begin{eqnarray}
\nonumber
\langle \hat K_x^2\rangle+\langle \hat K_y^2\rangle&=&-\frac{2\langle \hat K_z\rangle}{N}+\frac{\langle \hat N\hat K_z\rangle}{N}+\frac{\langle \hat N^2\hat K_z\rangle}{4N}\\
&&-\frac{\langle \hat N\hat K_z^2\rangle}{N}-\frac{4\langle \hat K_z^3\rangle}{N}+\frac{N^2}{16}+\frac{N}{4}.\label{MPcons}
\end{eqnarray}
We shall see later that this corresponds to a deformation of the familiar Bloch sphere of two-level systems to a teardrop shape on which the mean-field dynamics take place. 

\begin{figure}[tb]
\centering
\includegraphics[width=0.49\textwidth]{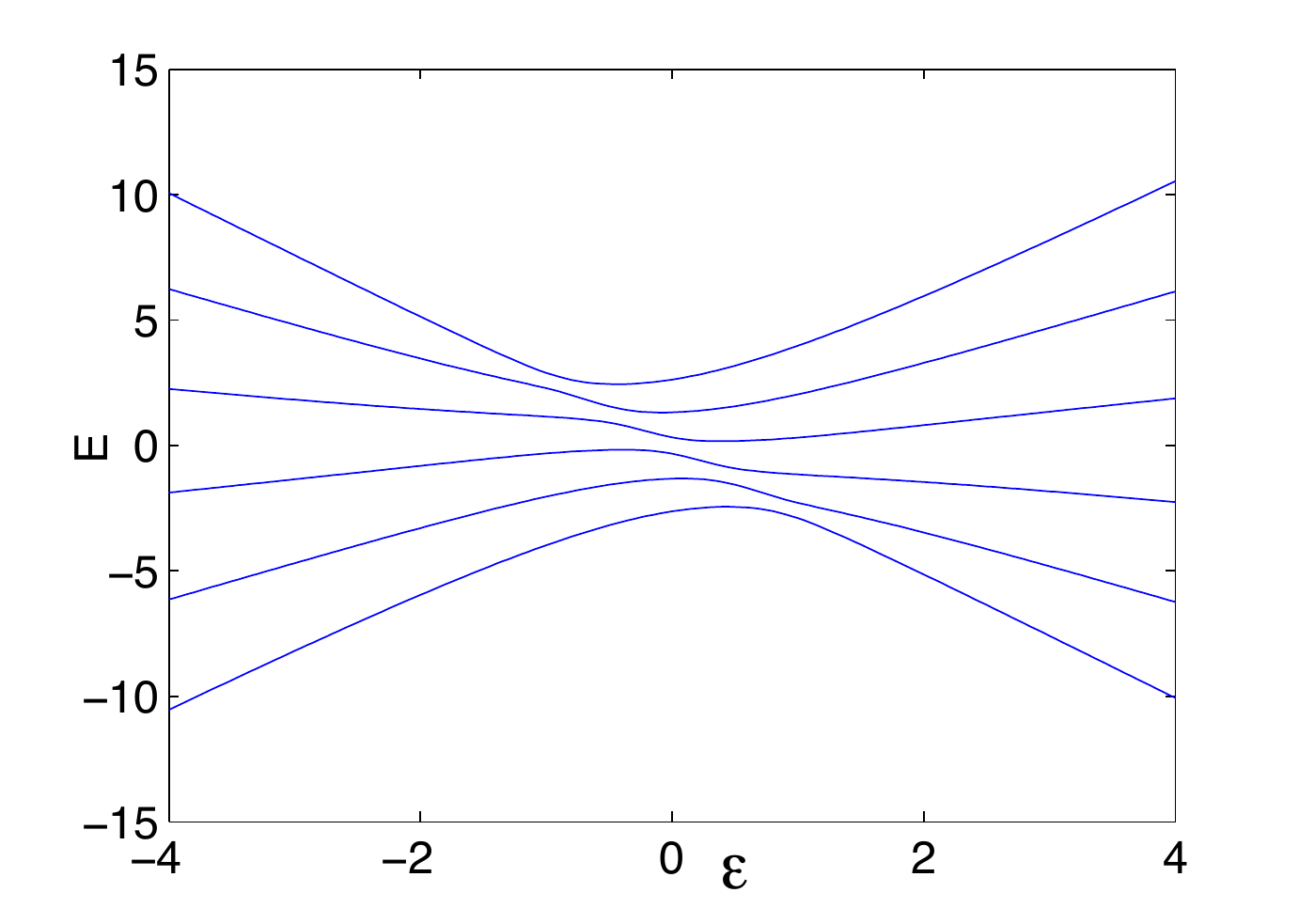} 
\includegraphics[width=0.49\textwidth]{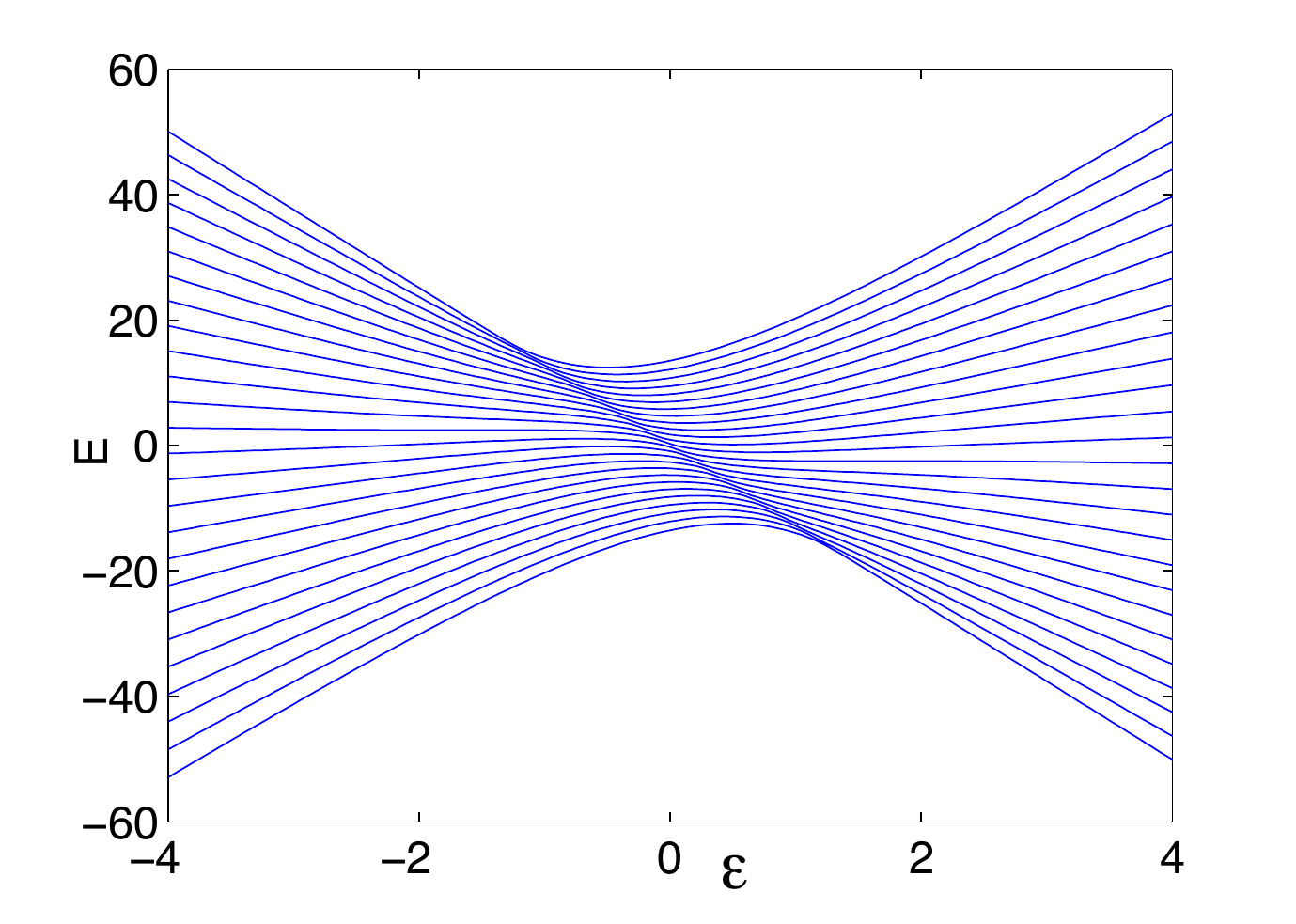}
\caption{(Color online) Many-particle spectrum in dependence on $\epsilon$ for $v=1$ and $N=10$ (top) and $N=50$ (bottom) particles.}
\label{fig_MPspec}
\end{figure}

From the commutation relation (\ref{eqn_defSU(2)_Kpm}) it follows that $\hat K_+$ and $\hat K_-$ are the usual lowering and raising operators for $\hat K_z$, from which one can deduce that the spectrum of $\hat K_z$ is equidistant \cite{Bona96}. For a given even particle number $N$ the eigenvalues of $\hat K_z$ run in integer steps from $-\frac{N}{4}$ to $\frac{N}{4}$.

While due to symmetry the operators $\hat K_x$ and $\hat K_y$ are isospectral, their spectrum differs from that of $\hat K_z$, and in particular, is not equidistant. The spectrum is depicted for the example $N=50$ in figure \ref{fig_Kxspec}. It can be seen that the eigenvalues are symmetric around zero, and approximately equidistant at the boundaries of the spectrum, while they are closer together around the centre. The eigenvalues can be obtained analytically using a Bethe ansatz approach \cite{Sant11}. 

Figure \ref{fig_MPspec} depicts the eigenvalues of the many-particle Hamiltonian (\ref{eqn_MPHam}) as a function of $\epsilon$ for $v=1$, and two different particle numbers. In comparison to the familiar Landau-Zener behaviour of a many-particle two-level system without atom-molecule conversion, the simple symmetry with respect to $\epsilon$ is lost, and we observe a cluster of narrow avoided crossings for intermediate values of $\epsilon$, stretching  from the upper end of the spectrum for negative values of $\epsilon$ to the lower for positive values. These are the reflection of the accumulation of eigenvalues in the spectrum of $\hat K_x$ in figure \ref{fig_Kxspec}. For large values of $\epsilon$ the many-particle spectrum is dominated by the equidistant spectrum of $\hat K_z$. The narrow avoided crossings lead to quasi-stationary states related to unstable stationary states in the mean-field description. These are related to the molecular mode, where all particles are paired up in diatomic molecules. 

The many-particle dynamics can be straight-forwardly obtained by integration of the Schr\"odinger equation on the $(\frac{N}{2}+1)$-dimensional Hilbert space. It is nevertheless instructive to study the Heisenberg equations of motion for the dynamical variables $\hat K_{x,y,z}$, given by
\begin{eqnarray}\nonumber
&\frac{\mathrm d}{\mathrm d t} \hat{K}_x & =-\epsilon \hat{K}_y  \\
&\frac{\mathrm d}{\mathrm d t} \hat{K}_y & =\epsilon \hat{K}_x +
\frac{v}{2}+ \frac{v}{8N}\left( \hat{N}^2 
 -8 \hat{K}_z\hat{N} - 
48 \hat{K}_z^2\right)\label{Kdyn}\\ 
&\frac{\mathrm d}{\mathrm d t} \hat{K}_z & = v\hat{K}_y.  \nonumber
\end{eqnarray}
Due to the nonlinearity of the commutator of $\hat K_x$ and $\hat K_y$ this is not a closed set of equations if $v\neq 0$, that is, the right hand side contains dynamical variables such as $\langle \hat K_z^2\rangle$ whose dynamics is not determined by the dynamical equations (\ref{Kdyn}). In the trivial case $v=0$, where the atomic and molecular mode decouple, we have a closed set of equations, describing rigid rotations around the $z$-axis. In the general case, taking the expectation value and neglecting quantum fluctuations, i.e., approximating expectation values of products with products of expectation values, yields the mean-field approximation we shall discuss in the following section. It is interesting to note that the same dynamical equations in terms of generators of a deformed $SU(2)$ algebra also appear in the context of fermionic atom-molecule conversion \cite{Tikh06b,Li09}.
\section{Mean-field approximation}
In this section we review the derivation of the mean-field dynamics, and summarise its most important features \cite{Sant06,Liu08,Li09,Liu10,Sant11,Khrip11,Li11b,Cui12,Shen13,Donl02}. Here we formulate the approximation in a way that is most natural from the perspective of a semiclassical limit. The mean-field approximation can be obtained in the limit of large Hilbert space dimension by replacing expectation values of products with products of expectation values in the dynamics of the operators $\hat K_j$ in equation (\ref{Kdyn}). Introducing the mean-field variables $s_j=\eta \langle \hat K_j\rangle$, with $\eta=\left(\frac{N}{2}+1\right)^{-1}\to 0$, the mean-field dynamical equations become
\begin{eqnarray}\nonumber
&\dot s_x& =-\epsilon s_y \\
&\dot s_y& =\epsilon s_x + \frac{v}{4}\left(1-4s_z-12s_z^2\right)\label{sdyn} \\ 
&\dot s_z& = v s_y\nonumber. 
\end{eqnarray}
These can be formulated as canonical Hamiltonian dynamics with the 
classical Hamiltonian function
\begin{equation}
\label{MF_Hamiltonian}
H=\eta\langle\hat H\rangle=\epsilon s_z+v s_x,
\end{equation}
and the Poisson brackets 
\begin{eqnarray}
\{s_x,s_y\}&=&\frac{1}{4}\left(1-4s_z-12s_z^2\right)\nonumber\\
\{s_y,s_z\}&=& -s_x\label{MF_Poisson}\\
\{s_z,s_x\}&=& -s_y\nonumber,
\end{eqnarray}
which directly follow from the many particle commutators with the identification 
\begin{equation}
\{s_j,s_k\}=\lim_{N\to \infty}i\eta\langle [\hat K_j,\hat K_k]\rangle.
\end{equation}

\begin{figure}
\centering
\includegraphics[width=0.26\textwidth]{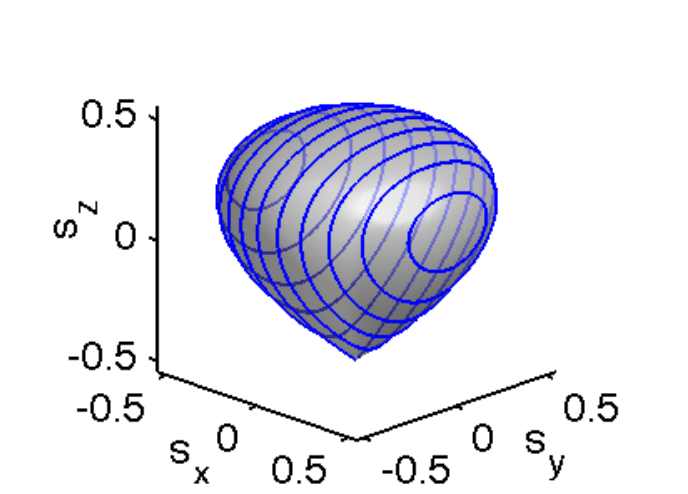} 
\includegraphics[width=0.21\textwidth]{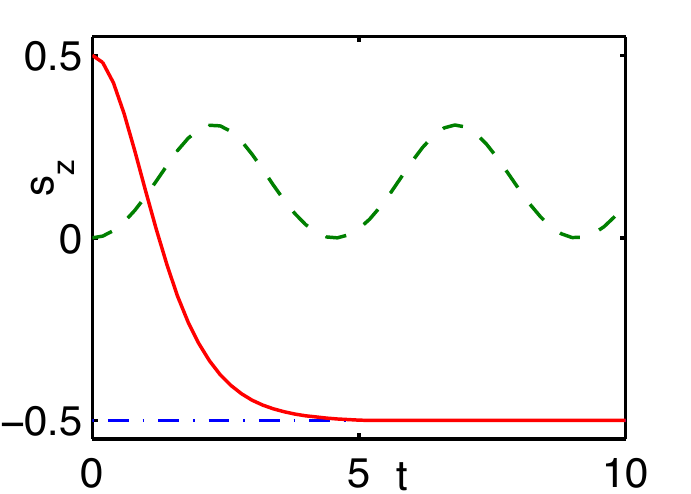}
\includegraphics[width=0.26\textwidth]{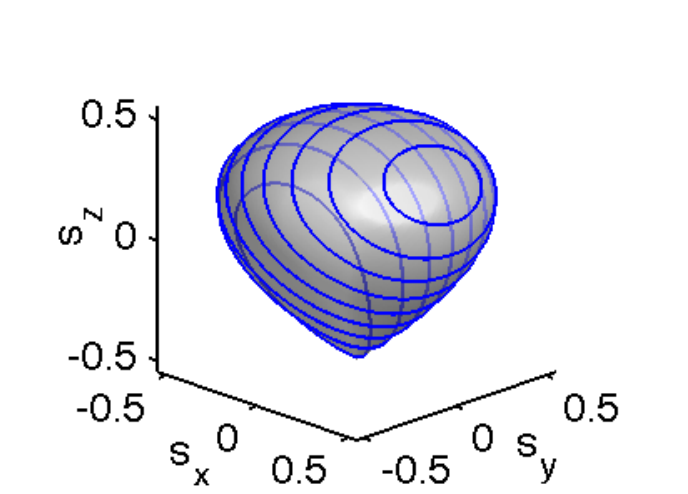}
\includegraphics[width=0.21\textwidth]{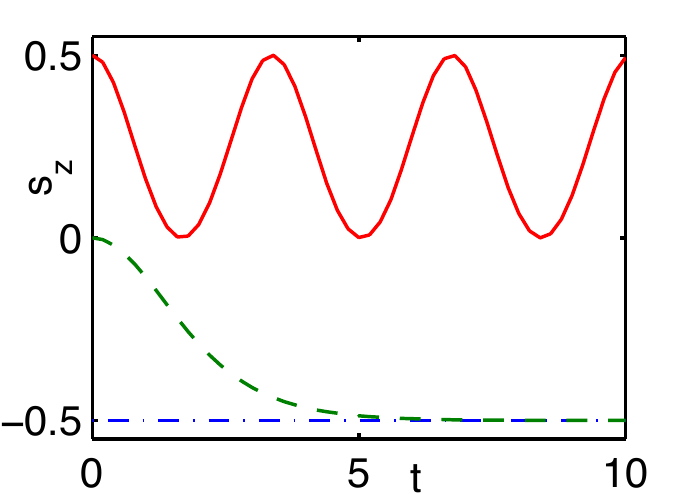}
\includegraphics[width=0.26\textwidth]{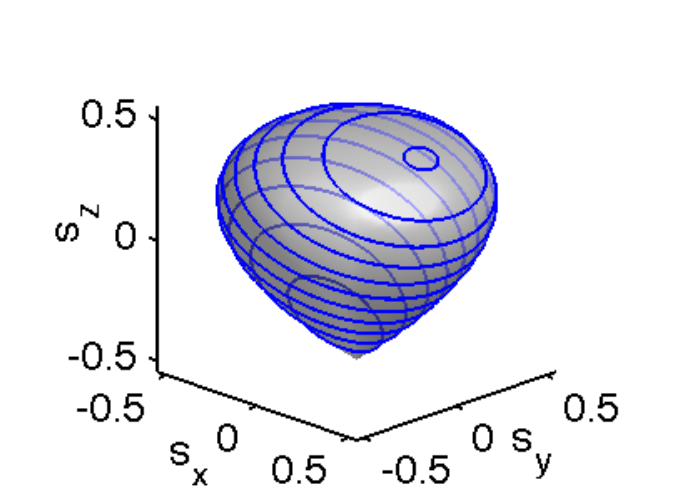}
\includegraphics[width=0.21\textwidth]{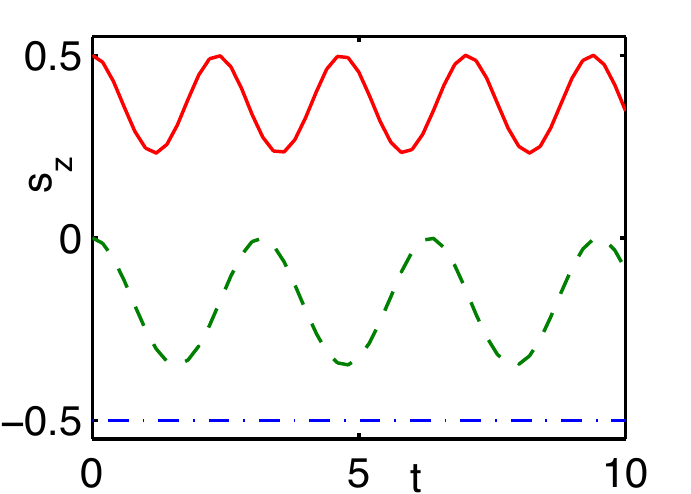}
\caption{(Color online) Mean-field dynamics on the deformed Bloch sphere for $v=1$ and $\epsilon=0,\,1,\,2$ (from top to bottom). The right panel shows the dynamics of $s_z$ for selected initial conditions.}
\label{fig_MFdyn}
\end{figure}

The dynamics (\ref{sdyn}) is confined to a two-dimensional surface given by the constraint
\begin{equation}
\label{MF_cons}
s_x^2+s_y^2=\frac{1}{4}(1-2s_z)(1+2s_z)^2=:r^2(s_z),
\end{equation}
with $s_z\in[-\frac{1}{2},\frac{1}{2}]$. This constraint also follows from the many-particle conserved quantity (\ref{MPcons}) in the mean-field limit. The resulting surface is depicted in figure \ref{fig_MFdyn}. It has a characteristic inverted teardrop shape, with a tip at $s_z=-\frac{1}{2}$. Note that similar shapes are known as Kummer shapes in the classical description of coupled oscillators of different frequencies \cite{Holm11,Kumm81,Kumm86}.

From the dynamical equations (\ref{sdyn}) it follows that all fixed points are located at $s_y=0$, and have to fulfil
\begin{equation}
\epsilon s_x=-\frac{v}{4}(\frac{1}{2}+s_z)(2-12s_z).
\end{equation}
Using the constraint (\ref{MF_cons}) this yields a polynomial in the $s_z$ component of the fixed points
\begin{equation}
\label{polyfp}
(\frac{1}{2}+s_z)^2\left[\frac{v^2}{4}-\epsilon^2-(3v^2-2\epsilon^2)s_z+9v^2s_z^2\right]=0.
\end{equation}
It can be verified that for each of the solutions $s_z\in[-\frac{1}{2},\frac{1}{2}]$ of (\ref{polyfp}) there is one corresponding value fo $s_x$ such that the dynamics is stationary. That is, there is always a fixed point at the tip of the teardrop, at $s_z=-\frac{1}{2}$. Depending on the values of $v$ and $\epsilon$ there can be one or two further fixed points. For values of $\epsilon$ that are smaller than a critical value $|\epsilon_{\rm crit}|=\sqrt{2}|v|$ there are three fixed points. A stability analysis reveals that the fixed point at the tip of the teardrop is a saddle point in this case, while the other two fixed points are elliptic. At the critical value $|\epsilon_{\rm crit}|=\sqrt{2}|v|$ one of the other two fixed points collides with the fixed point at the tip of the teardrop and then moves to unphysical values $s_z<-\frac{1}{2}$. The two fixed points interchange their stability in this transcritical bifurcation. That is, for values of $\epsilon$ larger than the critical value there are two elliptic fixed points, one of which is located at the tip, the other elsewehere on the teardrop. Examples of the dynamics for $v=1$ and different values of $\epsilon$ are shown in figure \ref{fig_MFdyn}. In the top two figures the fixed point at the tip of the teardrop is a saddle point, which is approached asymptotically by trajectories starting in a particular region on the teardrop, that shrinks as $\epsilon$ gets closer to the critical value. In the lowest figure the value of $\epsilon$ is super critical, and we observe the typical oscillations for trajectories in the neighbourhood of the elliptic fixed point at the tip of the teardrop. 

We can introduce a set of canonical variables $p$ and $q$ via the transformation
\begin{eqnarray}
s_x=r(p)\cos(q)\\
s_y=r(p)\sin(q)\\
s_z=p,
\end{eqnarray}
with $r(p)=\frac{\sqrt{(1-2p)(1+2p)^2}}{2}$, with $p\in[-\frac{1}{2},\frac{1}{2}]$ and $q\in[0,2\pi]$. It is straight forward to verify that this indeed recovers the Poisson brackets (\ref{MF_Poisson}) with the standard definition
\begin{equation}
\label{Poisson}
\{A,B\}:=\frac{\partial A}{\partial p}\frac{\partial B}{\partial q}-\frac{\partial A}{\partial q}\frac{\partial B}{\partial p}.
\end{equation}
In terms of $p$ and $q$ the dynamics is then given by the canonical equations $\dot q=\frac{\partial H}{\partial p}$ and $\dot p=-\frac{\partial H}{\partial q}$ with the Hamiltonian function (\ref{MF_Hamiltonian}) expressed in terms of $p$ and $q$:
\begin{equation}
H=\epsilon p+\frac{v}{2}\sqrt{(1-2p)(1+2p)^2}\cos(q).
\end{equation}

The mean-field dynamics can of course be expressed in terms of a nonlinear Schr\"odinger equation for an effective single-particle wave function. In the case of atom-molecule conversion there are two natural ways to define a mean-field wave function. The first, most commonly used, arises via the usual identification of the components of the mean-field wave function with the probability amplitudes to be in one of the states, in this case to be in the atomic or molecular state. That is, we make the identification $\hat a^{(\dagger)}\to \frac{1}{\sqrt{\eta}}\,\psi_a^{(*)}$, and $\hat b^{(\dagger)}\to\frac{1}{\sqrt{\eta}}\,\psi_b^{(*)}$. The resulting mean-field wave function is then normalised as $|\psi_a|^2+2|\psi_b|^2=2$, where $|\psi_a|^2/2$ is the probability to find the mean-field system in the atomic mode and $|\psi_b|^2$ the probability to find it in the molecular mode. 
The mean-field dynamics is then governed by the Hamiltonian dynamics 
\begin{equation}
{\rm i}\hbar \dot\psi_j=\frac{\partial H}{\partial \psi_j^*},
\end{equation}
with the Hamiltonian function $H$ expressed in terms of the $\psi_j$ as 
\begin{equation}
H=\frac{\epsilon}{4}\left(|\psi_a|^2-2|\psi_b|^2\right)+\frac{v}{2\sqrt{2}}\left(\psi_a^{*2}\psi_b+\psi_a^2\psi_b^*\right),
\end{equation}
that is 
\begin{equation}
{\rm i}\begin{pmatrix}\dot\psi_a\\ \dot\psi_b\end{pmatrix}=\begin{pmatrix}\frac{\epsilon}{4}& \frac{v}{\sqrt{2}}\psi_a^*\\ \frac{v}{2\sqrt{2}} \psi_a &-\frac{\epsilon}{2}\end{pmatrix}\begin{pmatrix}\psi_a\\ \psi_b\end{pmatrix}.
\end{equation}
With the identification 
\begin{equation}
\begin{split}s_{x}=\frac{1}{2\sqrt{2}}\left(\psi_{a}^{*2}\psi_{b}+\psi_{b}^{*}\psi_{a}^{2}\right),\\
s_{y}=\frac{1}{2\sqrt{2} \rm i}\left(\psi_{a}^{*2}\psi_{b}-\psi_{b}^{*}\psi_{a}^{2}\right),\\
s_{z}=\frac{1}{4}\left(\left|\psi_{a}\right|^{2}-2\left|\psi_{b}\right|^{2}\right).
\end{split}
\end{equation}
this yields the dynamical equations (\ref{sdyn}) as expected.

Alternatively, we can replace $\hat a^{2(\dagger)}\to \frac{1}{\eta}\,\chi_a^{(*)}$, and $\hat b^{(\dagger)}\to \frac{1}{\sqrt{\eta}}\,\chi_b^{(*)}$. Then we have $|\chi_a|+2|\chi_b|^2=2$, and the nonlinear Schr\"odinger equation follows from the many-particle dynamics as
\begin{equation}
{\rm i}\begin{pmatrix}\dot\chi_a\\ \dot\chi_b\end{pmatrix}=\begin{pmatrix}\frac{\epsilon}{2}& \sqrt{2}v|\chi_a|\\ \frac{v}{2\sqrt{2}}&-\frac{\epsilon}{2}\end{pmatrix}\begin{pmatrix}\chi_a\\ \chi_b\end{pmatrix}.
\end{equation}
In this version the probability to find the system in the atomic mode is given by $|\chi_a|/2$, while the probability to find the system in the molecular mode is $|\chi_b|^2$. The variables $s_j$ are then defined as 
\begin{equation}
\begin{split}s_{x}=\frac{1}{2\sqrt{2}}\left(\chi_{a}^{*}\chi_{b}+\chi_{b}^{*}\chi_{a}\right),\\
s_{y}=\frac{1}{2\sqrt{2}\rm i}\left(\chi_{a}^{*}\chi_{b}-\chi_{b}^{*}\chi_{a}\right),\\
s_{z}=\frac{1}{4}\left(\left|\chi_{a}\right|-2\left|\chi_{b}\right|^{2}\right).
\end{split}
\end{equation}
Their dynamics are again given by equation (\ref{sdyn}). 
\section{Mean-field and Many-particle correspondence}
\begin{figure}[tb]
\centering
\includegraphics[width=0.49\textwidth]{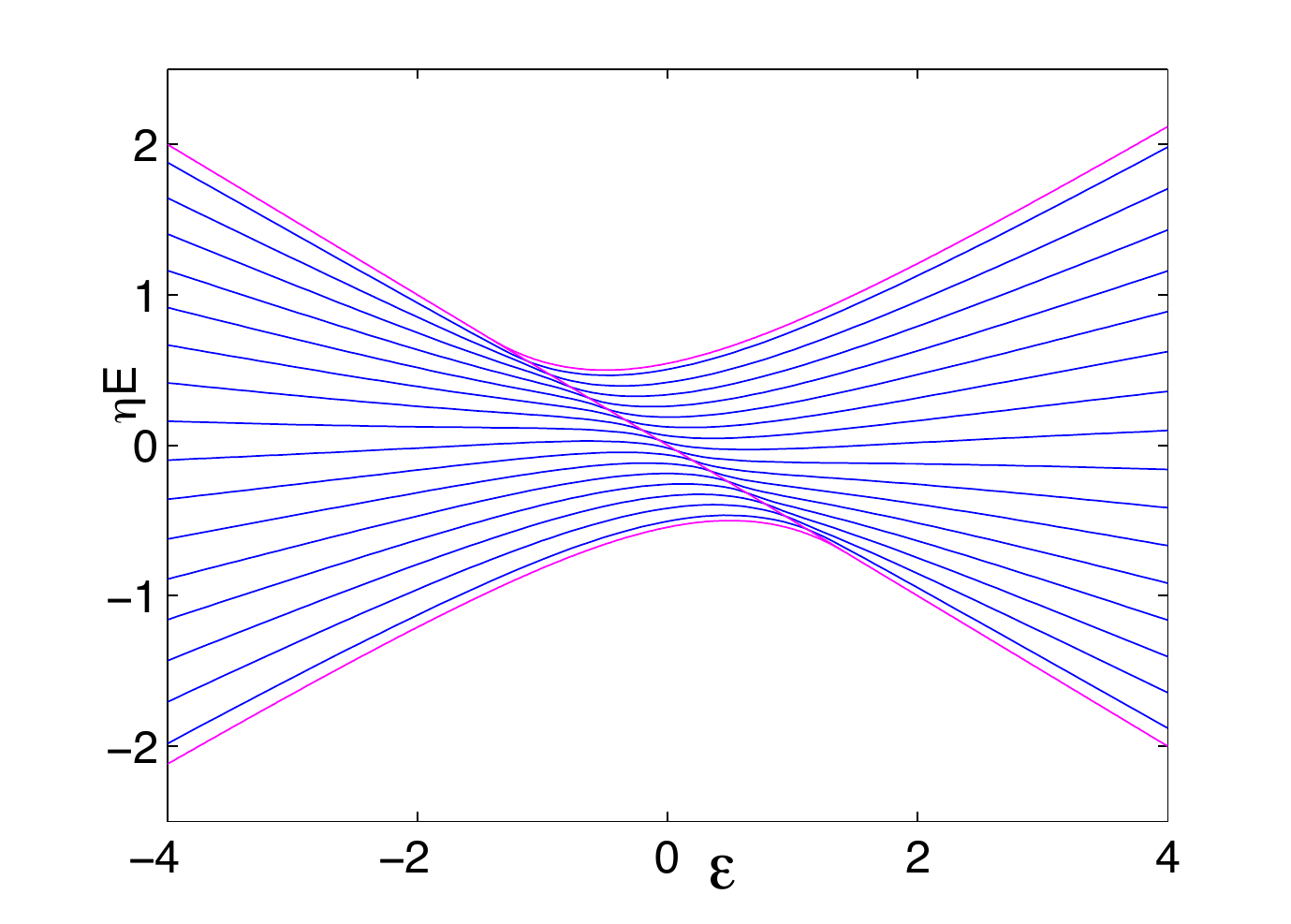}
\caption{(Color online) Many-particle (blue [dark gray]) and mean-field (magenta [light gray]) energies in dependence on $\epsilon$ for $v=1$ and $N=30$ particles.}
\label{fig_MF_MPspec}
\end{figure}
Let us first compare the spectral features of the mean-field and many-particle descriptions. For this purpose we define the mean-field energies as the values of the Hamiltonian function in the mean-field fixed points, that correspond to stationary solutions of the nonlinear Schr\"odinger equation. The resulting mean-field energies are plotted as a function of $\epsilon$ for $v=1$ in comparison with the many-particle energies for $N=30$ particles in figure \ref{fig_MF_MPspec}. It can be clearly seen how the pattern of narrow avoided crossings in the many-particle spectrum is closely following one of the mean-field energies. This energy corresponds to the unstable fixed point at the tip of the teardrop that is associated to the all-molecular mode. At the critical values of $\epsilon=\pm\sqrt{2}$ this becomes a stable elliptic fixed point, that is associated to the minimum and maximum eigenvalues of the many-particle system, respectively. The maximum and minimum mean-field energies bound the many-particle spectrum, if the latter is renormalised by the semiclassical parameter $\eta$, as is expected in a typical quantum-classical correspondence. This is very promising for a semiclassical quantisation that we shall attempt in the next section.

\begin{figure}[tb]
\centering
\includegraphics[width=0.26\textwidth]{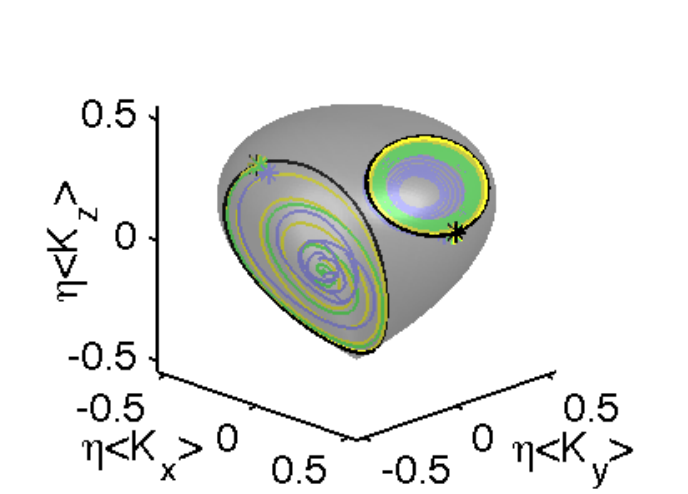} 
\includegraphics[width=0.21\textwidth]{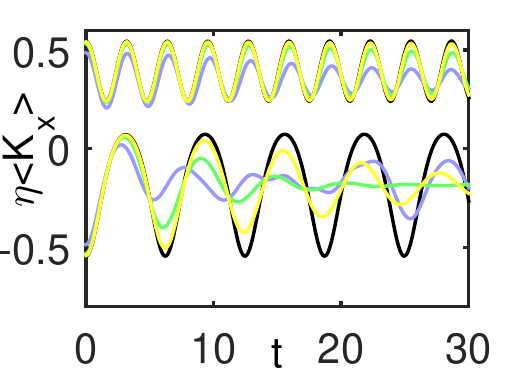}
\includegraphics[width=0.26\textwidth]{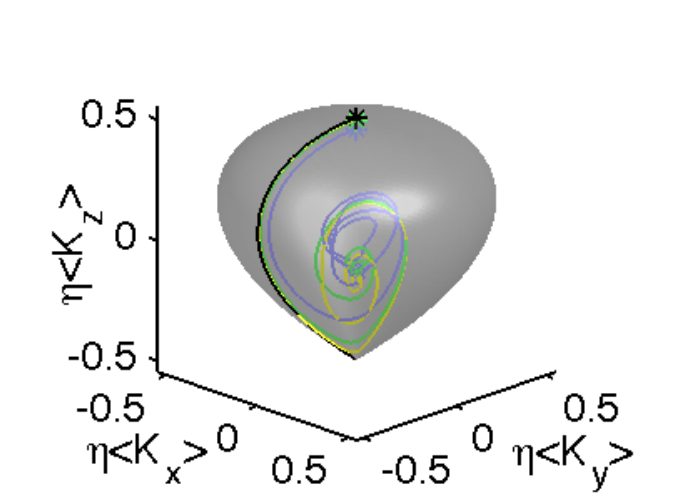}
\includegraphics[width=0.21\textwidth]{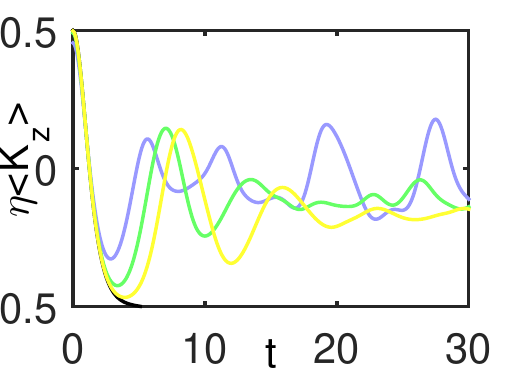}
\caption{(Color online) Mean-field (black) and many-particle dynamics for $v=1$ and $\epsilon=1$ (top) and $\epsilon=0$ (bottom), and different particle numbers. The blue (dark gray), green (gray), and yellow (light gray) curves correspond to $N=20,\,100,$ and $500$, respectively. The initial states are chosen as the ground states of $\pm\hat K_x$ in the top figure and $-\hat K_z$ in the bottom figure, and the initial conditions are marked by a star. The right figure on the top shows the dynamics of the $x$-component for $\epsilon=1$, and the figure on the bottom right shows the dynamics of the $z$-component for $\epsilon=0$.}
\label{fig_MP_MFdyn}
\end{figure}

\begin{figure}
\centering
\includegraphics[width=0.49\textwidth]{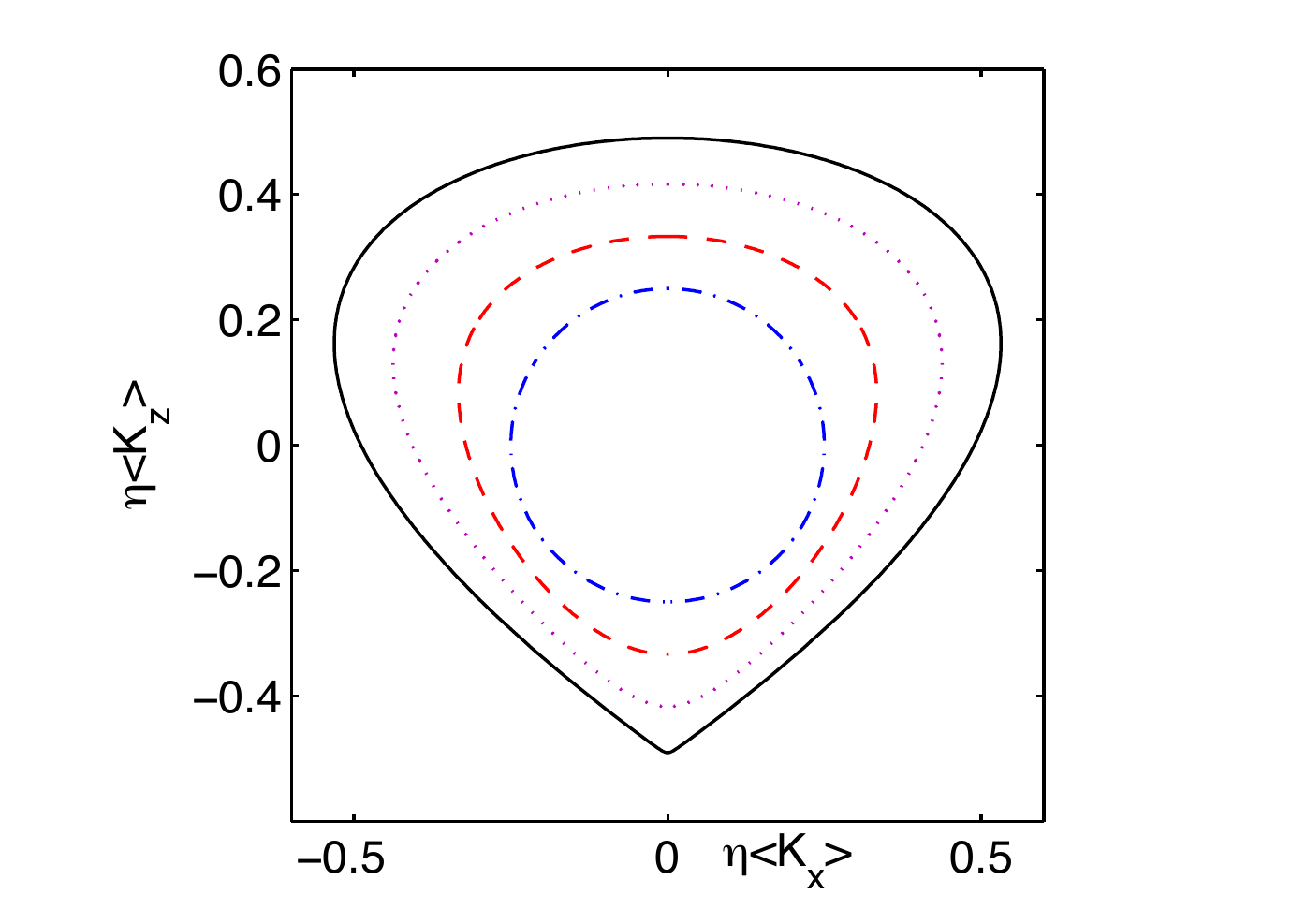}
\caption{(Color online) Expectation values of $\hat K_x$ and $\hat K_z$ for the ground states of the Hamiltonian (\ref{CS_Ham}) where $c=0$, $b$ varies from $-0.5$ to $0.5$ and $a=\pm\tfrac{1}{2}\sqrt{(1-2b)(1+2b)^2}$, for different particle numbers $N=2$ (blue dashed dotted line), $N=4$ (red dashed line), $N=10$ (magenta dotted line) and $N=100$ (black solid line). }
\label{fig_CS_surf}
\end{figure}

Let us now briefly turn to the correspondence between mean-field and many-particle dynamics. In figure \ref{fig_MP_MFdyn} we show several examples of mean-field trajectories (black lines) for different parameter values and initial conditions, in comparison to the corresponding many-particle dynamics. For comparison the initial many-particle state has been chosen as the ground state of a Hamiltonian of the type 
\begin{equation}
\label{CS_Ham}
\hat K=a\hat K_x+b\hat K_z+c\hat K_y,
\end{equation} 
whose expectation values of $\hat K_{x,y,z}$ lie as close as possible to the teardrop surface of the mean-field system. In figure \ref{fig_CS_surf} we show the expectation values of $\hat K_x$ and $\hat K_z$ for this family of states for $c=0$ and varying values of $a$ and $b$, for different particle numbers. It can be seen how the mean-field teardrop is approached with increasing particle number. See \cite{Li11b} for a comparison between these states and a new type of coherent states proposed there. 


In figure \ref{fig_MP_MFdyn} it can be seen how the many-particle dynamics closely follows the mean-field dynamics even for relatively small particle numbers in the vicinity of elliptic fixed points. Where the influence of the hyperbolic fixed point at the bottom of the teardrop is stronger, the deviations become larger and we observe the typical breakdown phenomenon of the many-particle dynamics. For longer times (not depicted), revival phenomena occur where the revival time rapidly increases with the particle number as expected \cite{Milb97}. The slow convergence of the many-particle dynamics towards the mean-field dynamics is particularly pronounced for those initial states whose mean-field dynamics asymptotically approaches the saddle point related to the all-molecular mode, as seen at the bottom in figure \ref{fig_MP_MFdyn}. The breakdown happens very rapidly here and even for longer times, no revivals are observed. This behaviour has been investigated in some detail in \cite{Vard01a}, and is typical for quantum-classical correspondence in the neighbourhood of hyperbolic fixed points. 

\section{Semiclassical quantisation}
In the following we shall investigate how semiclassical techniques might be modified to obtain an approximation for the many-particle eigenvalues from the mean-field dynamics. For this purpose it is convenient to use the approach by Braun  \cite{Brau93} that has been developed for general three-term recurrence relations, since the matrix representation of our many-particle Hamiltonian (\ref{eqn_MPHam}) in the eigenbasis of $\hat K_z$ is tri-diagonal, and thus the eigenvalue equation defines a three-term recurrence relation. The same procedure has been successfully applied to the spectrum of the two-mode Bose-Hubbard model in \cite{07semiMP,Niss10,Simo12}. Since the spectrum of $\hat K_z$ is related to that of two coupled harmonic oscillators, we expect that the semiclassical quantisation should yield exact results for the spectrum of $\hat K_z$ for arbitrary particle numbers. For this purpose the quantisation procedure of Braun has to be slightly modified. We shall give a brief overview over this modified version in what follows.

\begin{figure}
\centering
\includegraphics[width=0.23\textwidth]{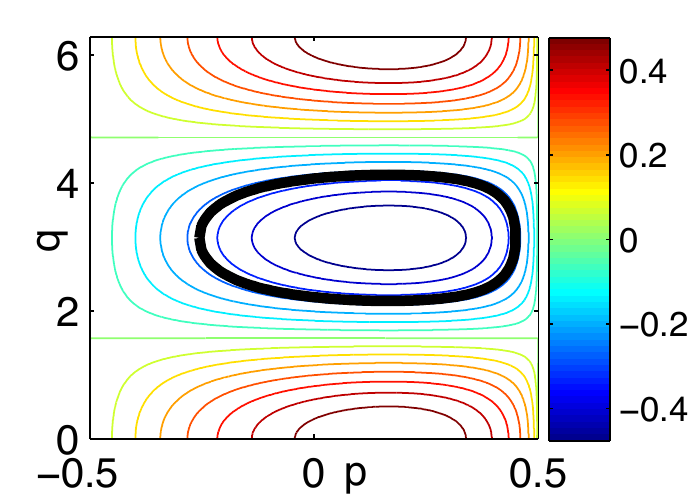}
\includegraphics[width=0.23\textwidth]{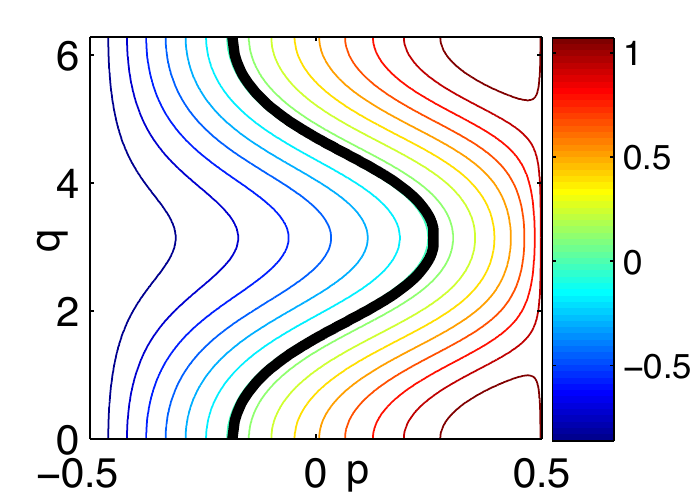}
\includegraphics[width=0.23\textwidth]{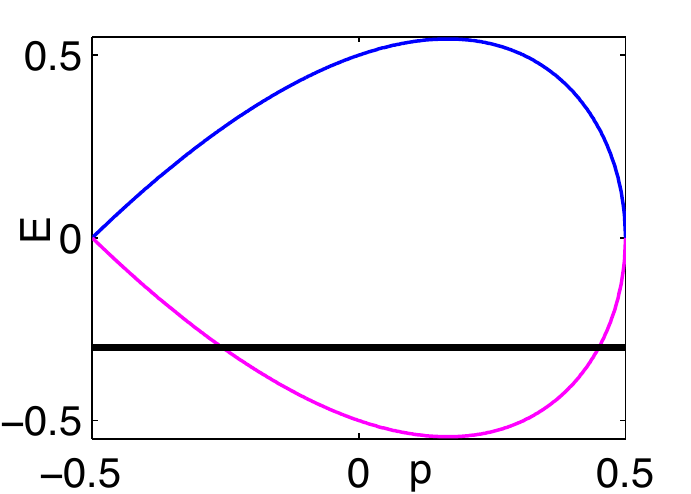}
\includegraphics[width=0.23\textwidth]{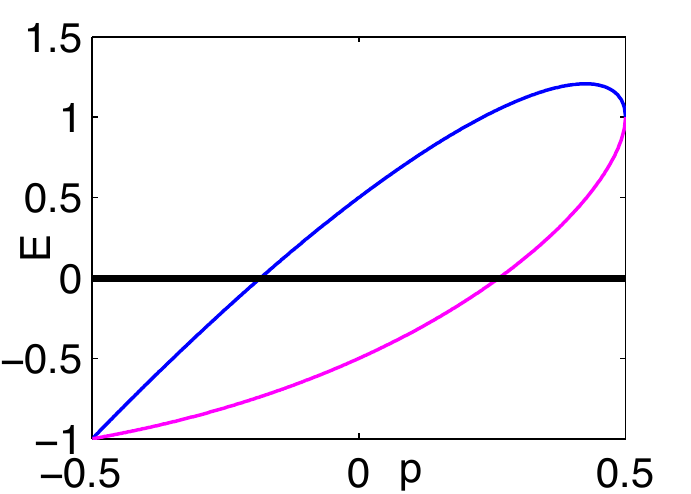}
\caption{(Color online) Mean-field phase space portraits (top) and potential curves (bottom) for $v=1$ and $\epsilon=0$ (left) and $\epsilon=2$ (right). The false colours in the phase space plots represent the energy values. An orbit belonging to one selected energy each (marked by a black line in the bottom pictures) is highlighted in each of the phase space plots by a thick black line.}
\label{fig_potential}
\end{figure}

The main idea is to derive a WKB type approximation for the wave function in dependence on $p$, the matching conditions between the allowed and forbidden regions then yield a Bohr-Sommerfeld type quantisation condition for the eigenvalues. The latter quantisation condition can in fact be motivated heuristically, by arguing that every many-particle state takes up on average a phase space volume of $h$, where the ground state and the highest exited state only take up the minimum uncertainty volume of $\frac{h}{2}$. However, the role of the semiclassical parameter $h$ in the present context is taken over by the inverse matrix size of the problem, that is, $\eta=\left(\frac{N}{2}+1\right)^{-1}$. The quantisation condition then reads
\begin{equation}
\label{eq:quantCondition}
S(\eta E_n)=2\pi\eta(n+\frac{1}{2}),
\end{equation}
where $S(E)$ is the mean-field phase-space area enclosed by the orbit of energy $E$.
To calculate these phase space areas it is convenient to introduce potential curves for $p$. These potential curves $U^{+}$ and $U^{-}$ are defined as the maximum and minimum functions of the Hamiltonian with respect to the angle dimension \cite{Brau93}. That is, we have 
\begin{equation} 
\label{pot.funs1}
U^{\pm}(p) =\epsilon p\pm\frac{v}{2}\sqrt{(1-2p)(1+2p)^2},
\end{equation}  
which join at $p=\pm\frac{1}{2}$, at the values $\pm\frac{\epsilon}{2}$. Two examples of the potential curves together with the corresponding phase-space portraits are shown in figure \ref{fig_potential}. The potential curves bound the possible values of the energy of the mean-field system for a given value of $p$. On the other hand, for a given value of the energy, the motion is restricted to values of $p$ where $U^-(p) \leq E \leq U^+(p)$, i.e. the classically allowed region. This region is bound by the classical turning points $p_{\pm}$, where $E = U^{\pm}(p_{\pm})$, which are found as the two roots of the polynomial 
\begin{equation}
\label{eq:turningpoints_pol}
2v^2p^3+(v^2+\epsilon^2)p^2-\left(\frac{v^2}{2}+2\epsilon E\right)p-\frac{v^2}{4}+E^2,
\end{equation}
that fall into the interval $[-\frac{1}{2},\frac{1}{2}]$. For any allowed value of the energies the polynomial has three real roots, two of which are located in the interval $[-\frac{1}{2},\frac{1}{2}]$. The third root, $p_0$, lies outside the physically relevant interval, to the left, that is, $p_0\leq-\frac{1}{2}\leq p_-$. The fixed points of the dynamics correspond to the extrema of the potential curves and the tip at which both potential curves meet at $p=-\frac{1}{2}$. For the extremal values of the energy the two turning points coalesce. 

\begin{figure}
\centering
\includegraphics[width=0.49\textwidth]{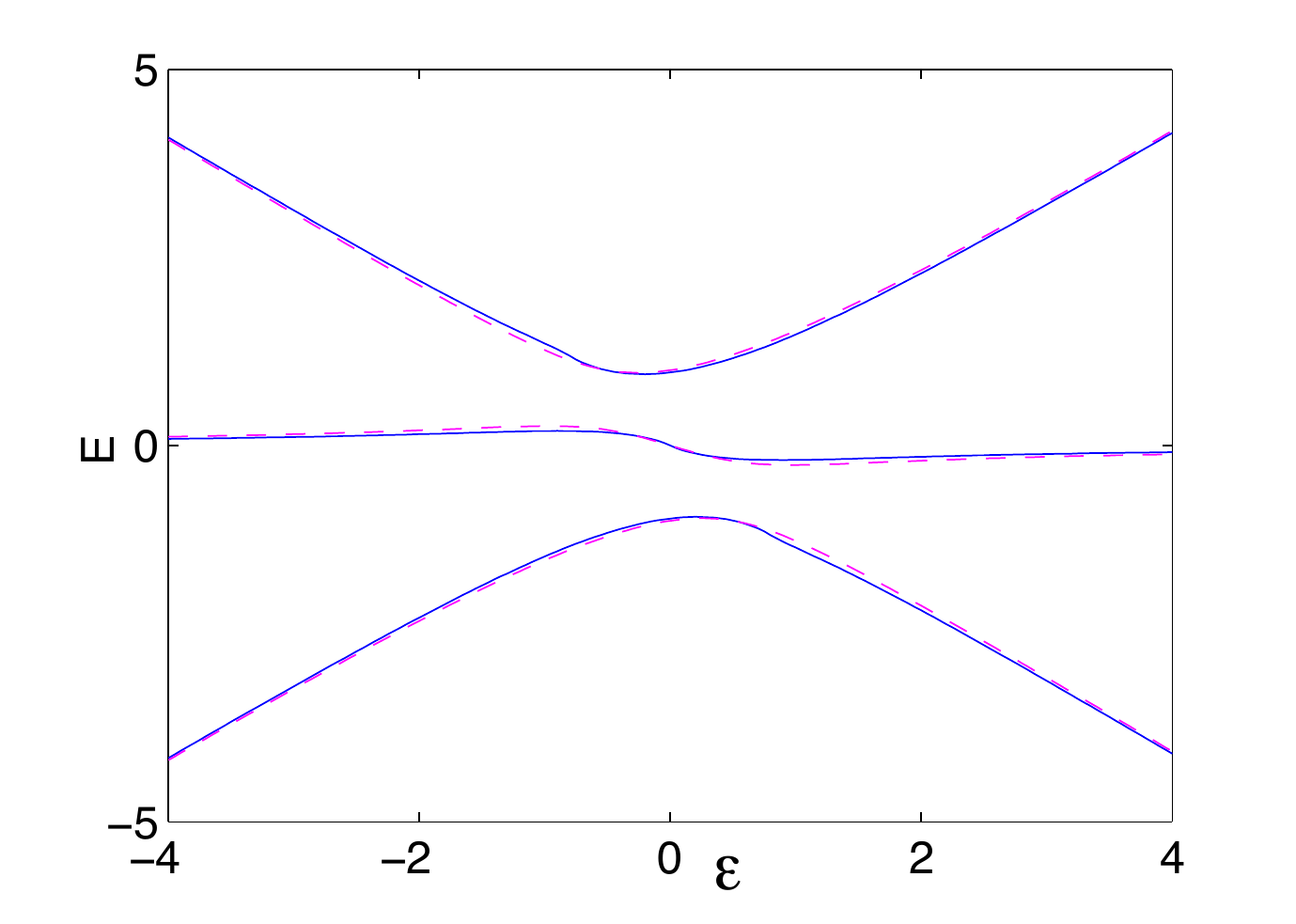}
\includegraphics[width=0.49\textwidth]{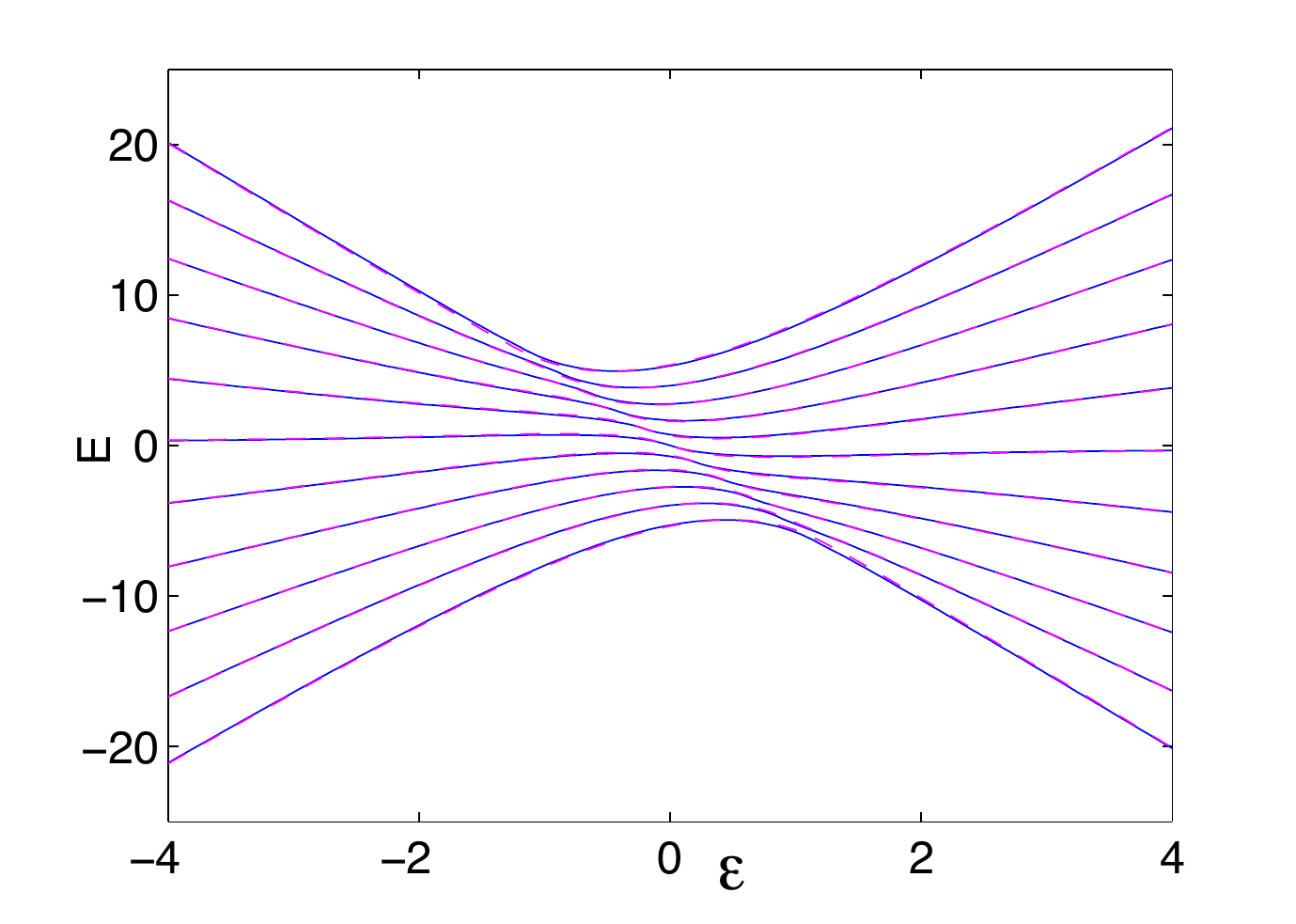}
\caption{(Color online) Many-particle (solid blue lines) and semiclassical (dashed magenta lines) eigenvalues in dependence on $\epsilon$ for $v=1$ and two different particle numbers. The upper panel corresponds to $N=4$ particles, the lower to $N=20$.}
\label{fig_quant_energy}
\end{figure}

The phase-space area enclosed by an orbit with energy $E$ can be written as the integral
\begin{equation}
\label{eqn_action_q(p)}
S(E) = \oint q(p,E) \mathrm{d}p,
\end{equation}
where $q(p)$ follows from the conservation of the energy as 
\begin{equation}
q(p)=\arccos\left(\frac{2(E-\epsilon p)}{\sqrt{(1-2p)(1+2p)^2}}\right),
\end{equation}
and care has to be taken that the area \textit{enclosed} by the curve is calculated, rather than the area outside the curve. Depending on which of the potential curves which turning point lies on,  
the enclosed area in equation (\ref{eqn_action_q(p)}) can be evaluated as 
\begin{equation}
S\left(E\right)=\begin{cases}
2\pi\left(p_+-p_{-}\right)-2\tilde S(E)\mbox{, } & p_{\pm}\mbox{ on }U_{-}\mbox{,}\\
2\pi\left(\frac{1}{2}-p_-\right)-2\tilde S(E)\mbox{, } & p_{-}\mbox{ on }U_{-},\mbox{ }p_{+}\mbox{ on }U_{+}\mbox{,}\\
2\pi\left(\frac{1}{2}+p_+\right)-2\tilde S(E) & p_{-}\mbox{ on }U_{+},\mbox{ }p_{+}\mbox{ on }U_{-}\mbox{,}\\
-2\pi+ 2\tilde S(E)& p_{\pm}\mbox{ on }U_{+},
\end{cases}
\end{equation}
with
\begin{equation}
\tilde S(E)=\int_{p_{-}}^{p_+}q\left(p\right)dp.
\end{equation}
In the case $v=0$ this can be evaluated analytically, and we obtain the exact eigenvalues for arbitrary particle numbers, as expected. For non-zero values of $v$ the quantisation condition (\ref{eq:quantCondition}) can be straight forwardly solved numerically to obtain the semiclassical many-particle spectrum from the mean-field dynamics. In figure \ref{fig_quant_energy} we show the resulting semi-classical eigenvalues in dependence on $\epsilon$ for $v=1$ and two different particle numbers, in comparison to the numerically exact many-particle energies. We observe that even for $N=4$ particles, corresponding to a relatively large value of $\eta=\frac{1}{3}$, the agreement is very good. As expected the quality of the semiclassical quantisation gets better with larger particle number and with larger $\epsilon$, as in the limit $v/\epsilon\to 0$ the quantisations yields exact results. 

The quantisation condition can be used to derive an 
analytic expression for the many-particle density of states in the semiclassical limit of 
large particle numbers \cite{Chil91,Aubr96}. Differentiating the quantisation condition \eqref{eq:quantCondition} with respect to the energy yields
\begin{equation}\label{eq:quant2}
\frac{\mathrm{d}S(E_n)}{\mathrm{d}E} = 2\pi\frac{\mathrm{d}n}{\mathrm{d}E}.
\end{equation}
That is, the many-particle density of states is given by the derivative of the mean-field phase space 
area with respect to the energy. This, on the other hand is given by the 
period of the mean-field orbit of the given energy \cite{Arno89}. That is, we have for the many-particle density of states
\begin{equation}\label{eq:ClassicalPeriod2}
\frac{\mathrm{d}n}{\mathrm{d}E}=\frac{T(E)}{2\pi},
\end{equation}
where $T(E)$ denotes the period of the orbit with energy $E$. 
This can be directly calculated from 
\begin{eqnarray}
T(E)=2\int_{p_{-}}^{p_{+}}\frac{dp}{\dot p}=2\int_{p_{-}}^{p_{+}}\left(\frac{\partial H}{\partial q}\right)^{-1}dp\nonumber\\
\label{period}
= 2\int_{p-}^{p+} \frac{dp}{\sqrt{(U^+ - E)(E-U^-)}}.
\end{eqnarray}
From the explicit form of the potential curves we find the period in terms of an 
elliptic integral as
\begin{equation}\label{eq:TofE_elliptic}
\begin{split}
T(E)&=\frac{2\sqrt{2}}{v\sqrt{p_+-p_0}}\, \int_{0}^{\frac{\pi}{2}} \frac{d\theta}{\sqrt{1-\frac{p_+-p_-}{p_+-p_0}\sin^2\theta}}\\
&=\frac{2\sqrt{2}}{v\sqrt{p_+-p_0}}\, {\rm K}\left(\frac{p_+-p_-}{p_+-p_0}\right),
\end{split}
\end{equation}
where $p_0$ is the third root of the polynomial (\ref{eq:turningpoints_pol}), and ${\rm K}$ denotes the complete elliptic integral of the first kind. The period diverges at the classical turning point, which can be seen as follows. At the value $E=-\frac{\epsilon}{2}$, corresponding to the orbit passing through the tip of the teardrop, the polynomial (\ref{eq:turningpoints_pol}) can be explicitly factorised as 
\begin{equation}
P(E=-\frac{\epsilon}{2})=(p+\frac{1}{2})^2(2v^2p-v^2+\epsilon^2).
\end{equation}
That is, in the subcritical case $|\epsilon|<\sqrt{2}|v|$, where the fixed point at the tip corresponds to a saddle point, we have $p_0=p_-=-\frac{1}{2}$, and $p_+=\frac{1}{2}-\frac{\epsilon^2}{2v^2}$, that is, we have $T(E)\propto K(1)$, which diverges. For supercritical values, on the other hand, where the fixed point corresponds to the minimum of the energy, we have $p_-=p_+=-\frac{1}{2}$, and $p_0=\frac{1}{2}-\frac{\epsilon^2}{2v^2}<-\frac{1}{2}$. Thus, the period has the finite value of $\frac{\sqrt{2}\pi}{\sqrt{\frac{\epsilon^2}{2}-v^2}}$ in this case.

\begin{figure}
\centering
\includegraphics[width=0.23\textwidth]{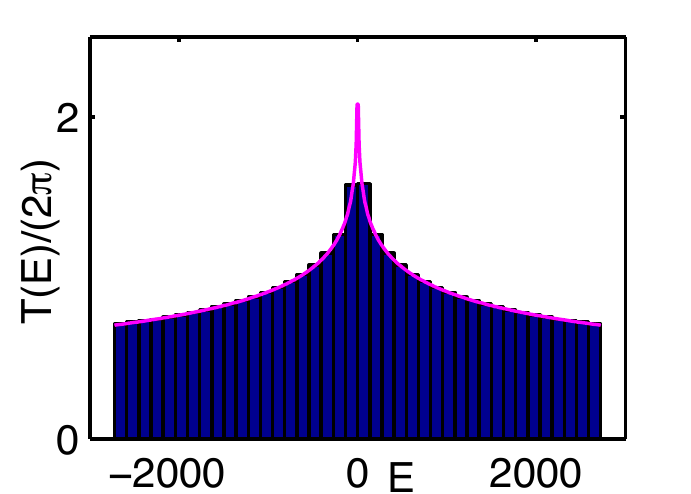}
\includegraphics[width=0.23\textwidth]{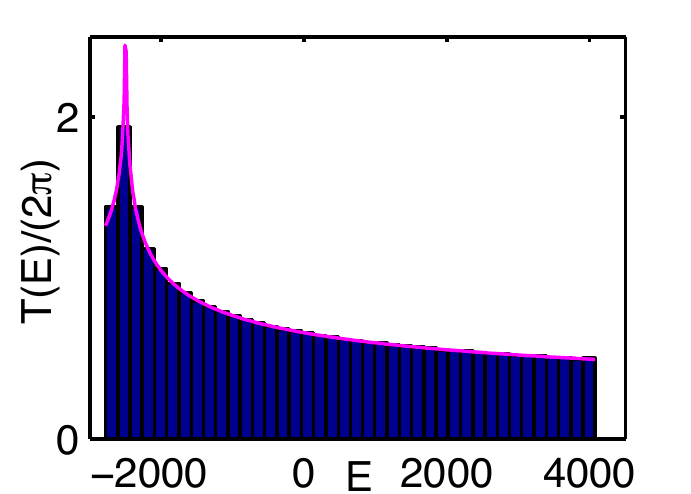}
\includegraphics[width=0.23\textwidth]{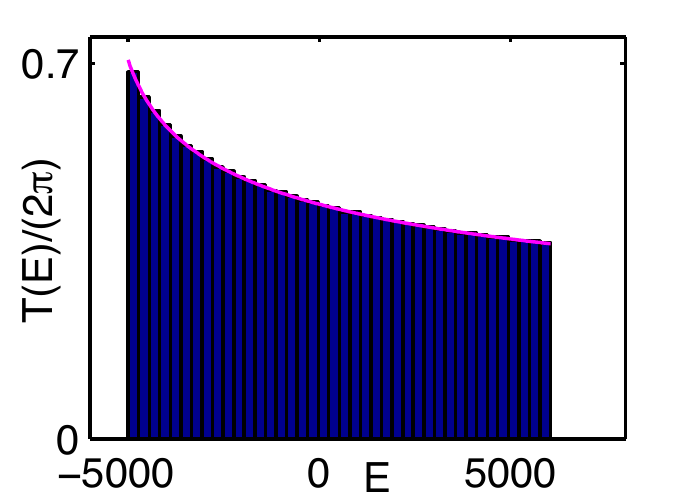}
\includegraphics[width=0.23\textwidth]{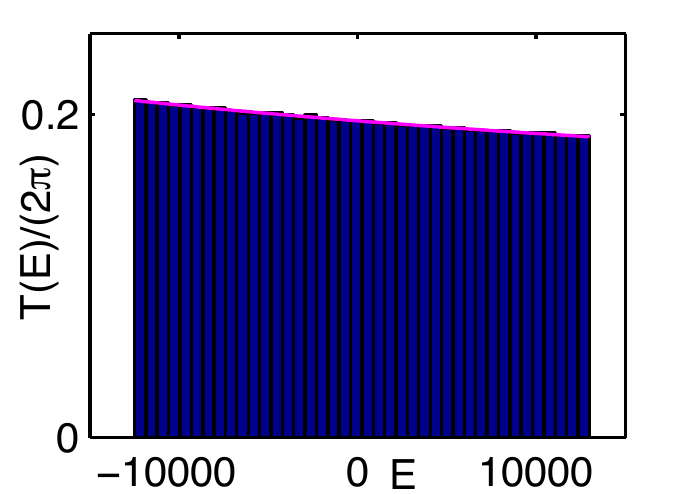}
\caption{(Color online) Density of states $\frac{\rm{d} n}{\rm{d} E}$ for the many-particle system in comparison with the mean-field periods for $v=1$ and different values of $\epsilon$. The many-particle density of states is approximated by the normalised histogram of the energies, for $N=10000$ particles. The mean-field result is depicted by the solid magenta line. The values of $\epsilon$ are $\epsilon=0,\,1,\,2,\,5$ from top left to bottom right.}
\label{fig_dos}
\end{figure}

Figure \ref{fig_dos} depicts the normalised histogram of many-particle eigenvalues for a large particle number of $N=10000$ in comparison to the mean-field periods (divided by $2\pi$), which are given by the analytical expression (\ref{eq:TofE_elliptic}). Note that the mean-field energies are rescaled with respect to the many-particle energies by $\eta$. An excellent agreement between the analytical expression and the many-particle density of states is observed. In particular the accumulation of states around the classical saddle point for values of $\epsilon$ below the critical value is nicely recovered in the many-particle histogram. Since the period of the orbit through the saddle point is infinite, the accumulation of many-particle eigenstates at this point leads to an actual divergence in the limit $N\to\infty$, which can be connected to a quantum phase transition \cite{Sant06}.

\begin{figure}
\centering
\includegraphics[width=0.49\textwidth]{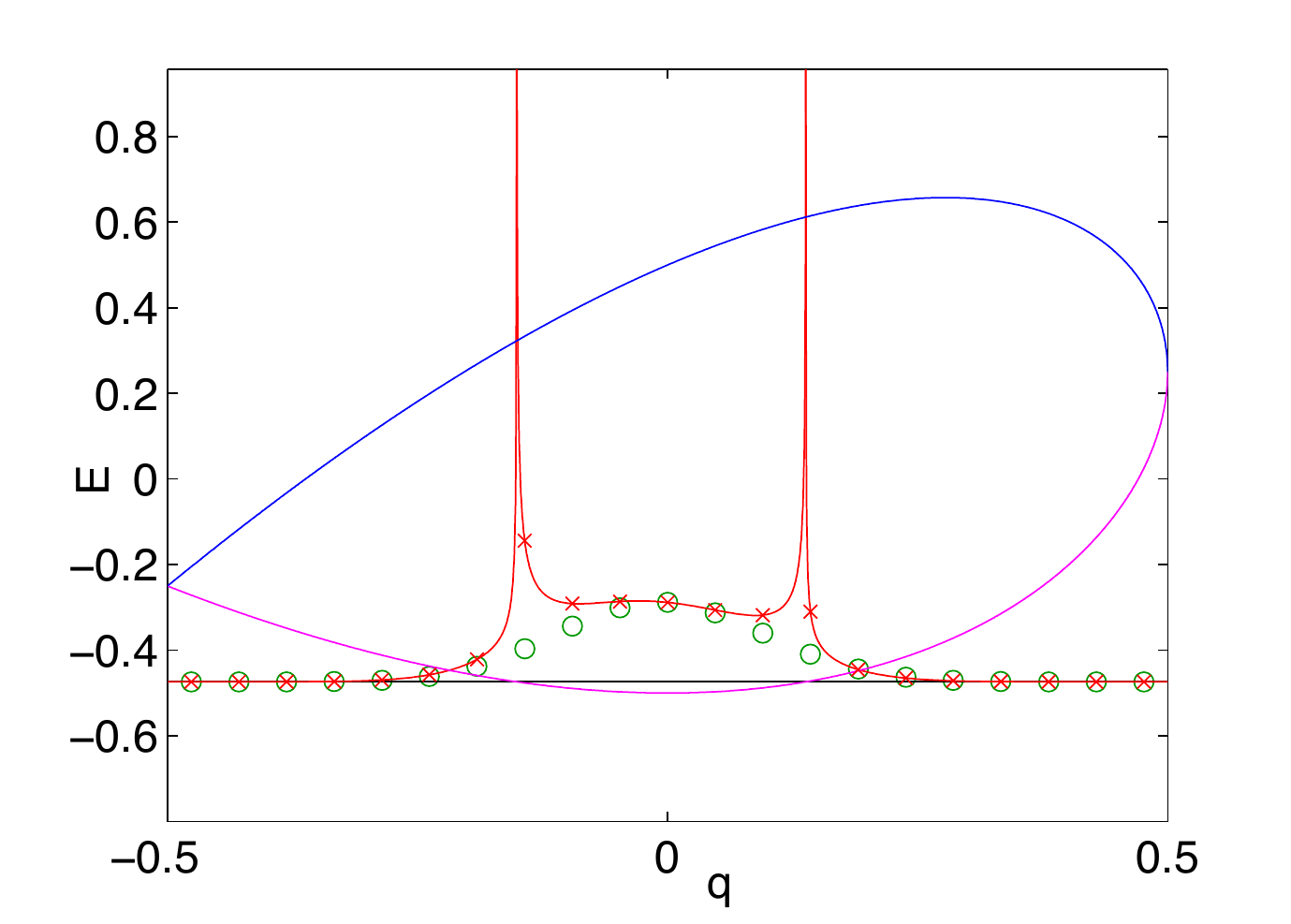}
\includegraphics[width=0.49\textwidth]{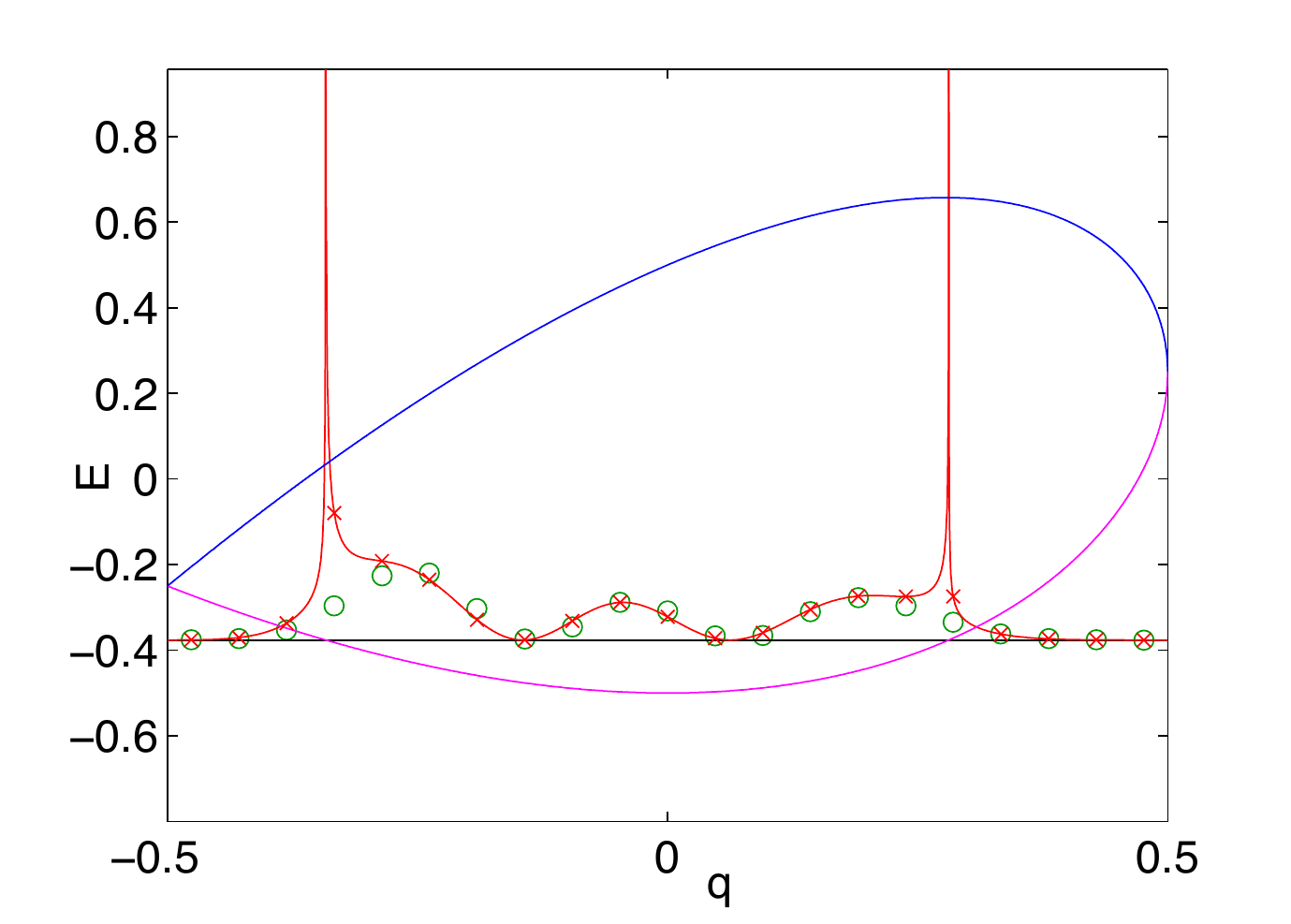}
\includegraphics[width=0.49\textwidth]{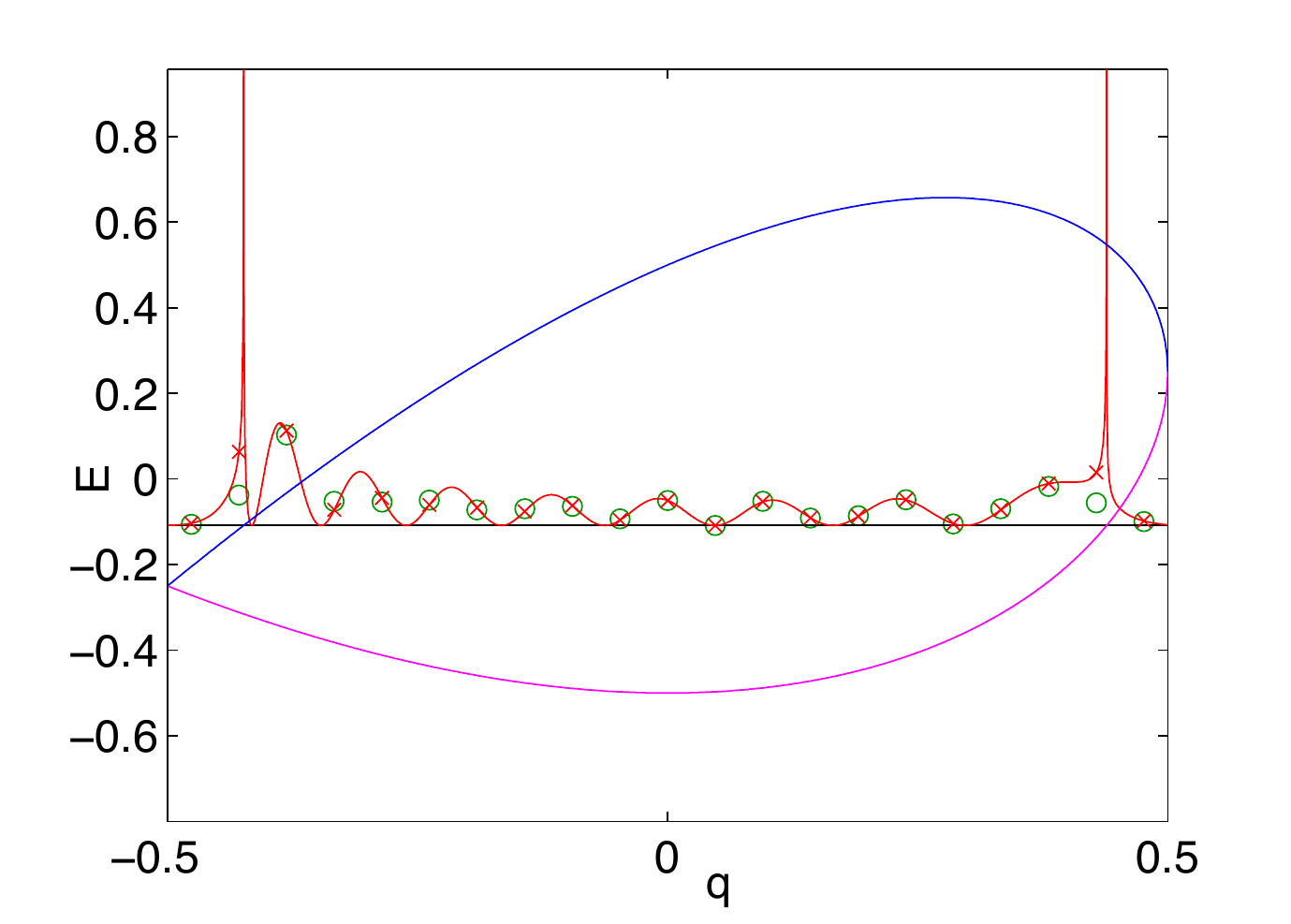}
\caption{(Color online) Exact $n$-th many-particle eigenvectors (green circles) with the WKB wave function (red) for $N=40$, $\epsilon=0.5$ and $v=1$ for  $n=1$ (top), $n=3$ (middle) and $n=10$ (bottom). The red crosses indicate the semiclassical wave function (\ref{eqn_semi_wfun_disc}). The solid red line, indicating the continuous semiclassical wave function according to (\ref{eqn_semi_wfun_cont}), is added to guide the eye.}
\label{fig_WKB_wavefun}
\end{figure}

We can also obtain an approximation for the components of the eigenvectors in the standard basis of $\hat K_z$, via a simple WKB ansatz. In the classically allowed region (between the two turning points) we make the ansatz 
\begin{equation}
\psi(p)=\sqrt{w_{cl}(p)}\left(A_+{\rm e}^{\frac{\rm i}{\eta}S\left(p\right)}+A_-{\rm e}^{-\frac{\rm i}{\eta}S\left(p\right)}\right),
\end{equation}
where $w_{cl}$ denotes the classical probability distribution
\begin{eqnarray}
\!\!\!\!\!\left|w_{cl}\left(p\right)\right|&=&\frac{1}{2T}\left(\frac{\partial H}{\partial q}\right)^{-1}\nonumber\\
&=&\frac{1}{2T\sqrt{\frac{v^{2}}{4}\left(1-2p\right)\left(1+2p\right)^{2}-\left(E-\epsilon p\right)^{2}}},
\end{eqnarray}
and the action $S(p)$ is given by
\begin{equation}
S\left(p\right)=\begin{cases}
\pi\left(p-p_{-}\right)-\tilde S(p)\mbox{, } & p_{\pm}\mbox{ on }U_{-}\mbox{,}\\
\pi\left(\frac{1}{2}-p\right)-\tilde S(p)\mbox{, } & p_{-}\mbox{ on }U_{-},\mbox{ }p_{+}\mbox{ on }U_{+}\mbox{,}\\
\pi\left(\frac{1}{2}+p\right)-\tilde S(p) & p_{-}\mbox{ on }U_{+},\mbox{ }p_{+}\mbox{ on }U_{-}\mbox{,}\\
-\pi+ \tilde S(p)& p_{\pm}\mbox{ on }U_{+},
\end{cases}
\end{equation}
with
\begin{equation}
\tilde S(p)=\int_{p_{-}}^{p}q\left(p\right)dp.
\end{equation}
In the forbidden region the WKB ansatz reduces to the single exponential decaying solution 
\begin{equation}
\left|\psi_{cl}\left(p\right)\right|^{2}=\frac{1}{2}\left|\omega_{cl}\left(p\right)\exp\left(-\frac{2i}{\eta}S\left(p\right)\right)\right|\mbox{,}
\end{equation}
with 
\begin{equation}
S\left(p\right)=\begin{cases}
\mp\int_{p}^{p_{-}}q\left(p\right)dp\mbox{,} & p<p_{-}\mbox{, }p_{-}\mbox{ on }U_{\pm}\mbox{,}\\
\mp\int_{p_{+}}^{p}q\left(p\right)dp\mbox{,} & p>p_{+}\mbox{, }p_{+}\mbox{ on }U_{\pm}\mbox{.}
\end{cases}
\end{equation}
The matching conditions at the boundary then impose the quantisation condition (\ref{eq:quantCondition}) and the absolute value of the WKB wave function in the classically allowed region becomes
\begin{equation}
\left|\psi_{cl}\left(p\right)\right|^{2}=2\left|w_{cl}\left(p\right)\right|\cos^{2}\left(\eta^{-1}S\left(p\right)-\frac{\pi}{4}\right). \label{eqn_semi_wfun_cont}
\end{equation}
While this expression appears to depend on the continuous variable $p$, for finite values of $\eta$ $p$ can only take on discrete values, due to the periodicity of its conjugate variable $q$, to which it can be related via a discrete Fourier transform. We thus have the semiclassical approximation 
\begin{equation}
|\Psi_n\rangle=\sum_{m=-\frac{N}{4}:1:\frac{N}{4}} \psi_{cl}(\eta m) |m\rangle, \label{eqn_semi_wfun_disc}
\end{equation}
where $|m\rangle$ denotes the eigenvectors of $\hat K_z$, that is, the states with $2m+\frac{N}{2}$ atoms and $\frac{N}{4}-m$ molecules. Figure \ref{fig_WKB_wavefun} depicts examples of the exact many-particle eigenvectors in comparison to the semiclassical approximation (normalised to fit the central maximum). As expected, the semiclassical approximation breaks down in the vicinity of the turning points, but approximates the many-particle wave functions well for other values. 

\section{Summary and Outlook}
While the role of many-particle effects in cold atom systems is crucial, in large realistic systems the mean-field approximation is often all that is accessible. Thus, the possibility to recover many-particle features from the mean-field description is an important addition to the theoretical toolbox for cold atoms. Here we have demonstrated that semiclassical methods can be modified to deduce many-particle properties for atom-molecule conversion systems from the mean-field dynamics alone. We have considered the eigenvalues and eigenvectors here; an extension to dynamical properties via semiclassical propagators is an important topic for future investigations. A non-trivial issue is the generalisation to more realistic models with many modes for both atoms and molecules. Progress has been made in this direction for cold atomic systems without atom-molecule conversion in \cite{Simo14,Visc11}. The combination with the results obtained here suggests that this goal is not out of reach. An obstacle for the application of some semiclassical techniques is the absence of a well-defined set of condensed states that coincide with coherent states of the deformed $SU(M)$ algebra for atom-molecule conversion systems. There have been proposals for coherent states in \cite{Suni00,Li11b}, however they do not fulfil all the properties that one relies on in the case of cold atoms in $M$ mode systems when employing $SU(M)$ coherent states \cite{07phase}. The connection between coherent states and projective manifolds \cite{10veronese} for $SU(M)$ systems might lead a way forward. 

\section*{Acknowledgements}
The authors thank Hans J\"urgen Korsch for valuable comments on the manuscript. EMG acknowledges support via the Imperial College JRF scheme, the L'Or\'eal UNESCO Women in Science programme, and from the Royal Society. AR acknowledges support from an EPSRC DTA grant.

\end{document}